\documentclass[journal]{IEEEtran}
\ifCLASSINFOpdf
\else
\fi

\usepackage{amssymb}
\usepackage{latexsym}
\usepackage{stmaryrd}
\usepackage{graphicx}

\begin{document}
%
\title{Deterministic Construction of Compressed Sensing Matrices using BCH Codes}
%
%
%

\author{Arash~Amini,
        and~Farokh~Marvasti,~\IEEEmembership{Senior~Member,~IEEE}
\thanks{A. Amini and F. Marvasti are with the Department of Electrical Engineering, Advanced Communication Research Institute (ACRI), Sharif University of Technology, Tehran,
Iran e-mail: arashsil@ee.sharif.edu , marvasti@sharif.edu.}
\thanks{Manuscript received July ?, 2009}
}

\maketitle

\begin{abstract}
In this paper we introduce deterministic $m\times n$ RIP fulfilling $\pm 1$ matrices of order $k$ such that $\frac{\log m}{\log k}\approx \frac{\log(\log_2 n)}{\log(\log_2 k)}$. The columns of these matrices are binary BCH code vectors that their zeros are replaced with $-1$ (excluding the normalization factor). The samples obtained by these matrices can be easily converted to the original sparse signal; more precisely, for the noiseless samples, the simple Matching Pursuit technique, even with less than the common computational complexity, exactly reconstructs the sparse signal. In addition, using Devore's binary matrices, we expand the binary scheme to matrices with $\{0,1,-1\}$ elements.
\end{abstract}


%
\IEEEpeerreviewmaketitle

\section{Introduction}
%
%
%
%
\IEEEPARstart{D}{ecreasing} the number of required samples for unique representation of a class of signals known as \emph{sparse} has been the subject of extensive research in the past five years. The field of compressed sensing which was first introduced in \cite{Donoho2006} and further in \cite{Candes2006a,Candes2006b}, deals with reconstruction of a $n\times 1$ but $k$-sparse vector $\mathbf{x}_{n\times 1}$ from its linear projections ($\mathbf{y}_{m\times 1}$) onto an $m$-dimensional ($m\ll n$) space: $\mathbf{y}_{m\times 1}=\mathbf{\Phi}_{m\times n}\mathbf{x}_{n\times 1}$. The two main concerns in compressed sensing are 1) selecting the sampling matrix $\mathbf{\Phi}_{m\times n}$ and 2) reconstruction of $\mathbf{x}_{n\times 1}$ from the measurements $\mathbf{y}_{m\times 1}$ by exploiting the sparsity constraint.

In general, the exact solution to the second problem, is shown to be an NP-complete problem \cite{Candes2005}; however, if the number of samples ($m$) exceeds the lower bound of $m >\mathcal{O}\big(k\log(n/k)\big)$, $\ell_1$ minimization (Basis Pursuit) can be performed instead of the exact $\ell_0$ minimization (sparsity constraint) with the same solution for almost all the possible inputs \cite{Candes2005,Candes2006a}. There are also other techniques such as greedy methods \cite{Tropp2007,Needell2008} that can be used.

The first problem (sampling matrix) is usually treated by random selection of the matrix; among the well-known random matrices are i.i.d Gaussian \cite{Donoho2006} and Rademacher \cite{Baraniuk2006} matrices. Before addressing some of the deterministic matrix constructions, we first describe the well known Restricted Isometry Property (RIP) \cite{Candes2006a}:

We say that the matrix $\mathbf{A}_{m\times n}$ obeys RIP of order $k$ with constant $0\leq\delta_k<1$ (RIC) if for all $k$-sparse vectors $\mathbf{x}_{n\times 1}$ we have:
\begin{eqnarray}
1-\delta_k\leq\frac{\|\mathbf{A}\mathbf{x}\|^2_{\ell_2}}{\|\mathbf{x}\|^2_{\ell_2}}\leq 1+\delta_k
\end{eqnarray}

In other words, RIP of order $k$ implies that each $k$ columns of the matrix $\mathbf{A}$ resembles a quasi-orthonormal set: if $\mathbf{B}_{m\times k}$ is formed by $k$ different columns of $\mathbf{A}$, all the eigenvalues of the Grammian matrix $\mathbf{B}^T\mathbf{B}$ should lie inside the interval $[1-\delta_k~,~1+\delta_k]$.

RIP is a sufficient condition for stable recovery. The basis pursuit and greedy methods can be applied for recovery of $k$-sparse vectors from noisy samples with good results if the matrix $\mathbf{A}$ obeys RIP of order $2k$ with a good enough constant $\delta_{2k}$ \cite{Candes2008,Needell2008}.

In this paper we are interested in deterministic as opposed to random sampling matrices. Deterministic sampling matrices are useful because in practice, the sampler should finally choose a deterministic matrix; realizations of the random matrices are not guaranteed to work. Moreover, by proper choice of the matrix, complexity or compression rate may be improved. In deterministic sampling matrix, we are looking for $m\times n$ matrices which obey the RIP of order $k$. It is well-known that any $k$ columns of a $k\times n$ Vandermond matrix are linearly independent; thus, if we normalize the columns, for all values of $n$, the new matrix satisfies the RIP condition of order $k$. In other words, arbitrary RIP-constrained matrices could be constructed in this way; however, when $n$ increases, the constant $\delta_k$ rapidly approaches $1$ and some of the $k\times k$ submatrices become ill-conditioned \cite{Cohen2009} which makes the matrix impractical. In \cite{DeVore2007}, $p^2\times p^{r+1}$ matrices ($p$ is a power of a prime integer) with $0,1$ elements (prior to normalization) are proposed which obey RIP of order $k$ where $kr<p$. Another binary matrix construction with $m=k2^{\mathcal{O}(\log\log n)^E}$ measurements ($E>1$) is investigated in \cite{Indyk2008conf} which employs hash functions and extractor graphs. The connection between coding theory and compressed sensing matrices is established in \cite{Howard2008conf} where second order Reed-Muller codes are used to construct $2^{l}\times 2^{\frac{l(l+1)}{2}}$ matrices with $\pm 1$ elements; unfortunately, the matrix does not satisfy RIP for all k-sparse vectors. Complex $m^2\times m$ matrices with chirp-type columns are also conjectured to obey RIP of some order \cite{Applebauma2009}. Recently, almost bound-achieving matrices have been proposed in \cite{Calderbank2009} which, rather than the exact RIP, satisfy statistical RIP (high probability that RIP holds). In this paper, we explicitly introduce $(2^l-1)\times 2^{2^{\mathcal{O}(\frac{l}{j}\log j)}}$ matrices with $\pm 1$ elements which obey the exact RIP for $k<2^j$. The new construction is achieved by replacing the zeros of the linear binary block codes (specially BCH codes) by $-1$. In this approach, we require binary codes with minimum distances as large as almost half their code length; the existence of these codes will be shown by providing  BCH codes. 

The rest of the paper is organized as follows: In the next section we show the connection between linear block codes and construction of RIP-fulfilling $\pm 1$ matrices. In section \ref{sec:dminCodes} we introduce BCH codes that meet the requirements to produce compressed sensing matrices. Matrix construction and recovery of the sparse signal from the samples using the matching pursuit method is discussed in section \ref{sec:SampRec}. The introduced matrices are combined with a previous scheme to form $0,\pm 1$ matrices in section \ref{sec:ThreeElement} and finally, section \ref{sec:Conclusion} concludes the paper.


\section{Matrix Construction via Linear Codes}\label{sec:CSCoding}

In this section we will describe the connection between the sampling matrix and coding theory. Since the parameters $k,n$ are used in both compressed sensing and coding theory, we distinguish the two by using the $\tilde{_{\Box}}$ notation for coding parameters; i.e., $\tilde{n}$ refers to the code length while $n$ denotes the number of columns of the sampling matrix. 

Let $\mathcal{C}(\tilde{n},\tilde{k};2)$ be a linear binary block code and $\mathbf{1}_{\tilde{n}\times 1}$ be the all $1$ vector. We say $\mathcal{C}$ is 'symmetric' if $\mathbf{1}_{\tilde{n}\times 1}\in\mathcal{C}$. For symmetric codes, if $\mathbf{a}_{n\times 1}$ is a code vector, due to the linearity of the code, complement of $\mathbf{a}_{n\times 1}$ which is defined as $\mathbf{a}_{n\times 1} \oplus \mathbf{1}_{\tilde{n}\times 1}$ is also a valid code vector; therefore, code vectors consist of complement couples.
\\
\newtheorem{theo}{\textbf{Theorem}}
\begin{theo}\label{theo:CodingCS}
Let $\mathcal{C}(\tilde{n},\tilde{k};2)$ be a symmetric code with the minimum distance $\tilde{d}_{min}$ and let $\tilde{\mathbf{A}}_{\tilde{n}\times 2^{\tilde{k}-1}}$ be the matrix composed of code vectors as its columns such that from each complement couple, exactly one is selected. Define:
\begin{eqnarray}
\mathbf{A}_{\tilde{n}\times 2^{\tilde{k}-1}}\triangleq \frac{1}{\sqrt{\tilde{n}}}\bigg(2\tilde{\mathbf{A}}_{\tilde{n}\times 2^{\tilde{k}-1}} - \big(1\big)_{\tilde{n}\times 2^{\tilde{k}-1}}\bigg)
\end{eqnarray}
Then, $\mathbf{A}$ satisfies RIP with the constant $\delta_k=(k-1)\big(1-2\frac{\tilde{d}_{min}}{\tilde{n}}\big)$ for $k<\frac{\tilde{n}}{\tilde{n}-2\tilde{d}_{min}}+1$ ($k$ is the RIP order).
\end{theo}

\vspace{0.5cm}

\textbf{Proof}. First note that the columns of $\mathbf{A}$ are normal. In fact $2\tilde{\mathbf{A}}_{\tilde{n}\times 2^{\tilde{k}-1}} - \big(1\big)_{\tilde{n}\times 2^{\tilde{k}-1}}$ is the same matrix as $\tilde{\mathbf{A}}$ where zeros are replaced by $-1$; hence, absolute value of each element of $\tilde{\mathbf{A}}$ is equal to $\frac{1}{\sqrt{\tilde{n}}}$ which reveals that the columns are normal.

To prove the RIP, we use a similar approach to that of \cite{DeVore2007}; we show that for each two columns of $\mathbf{A}$, the absolute value of their inner product is less than $\frac{\tilde{n}-2\tilde{d}_{min}}{\tilde{n}}$. Let $\mathbf{a}_{\tilde{n}\times 1}, \mathbf{b}_{\tilde{n}\times 1}$ be two different columns of $\mathbf{A}$ and $\tilde{\mathbf{a}}_{\tilde{n}\times 1}, \tilde{\mathbf{b}}_{\tilde{n}\times 1}$ be their corresponding columns in $\tilde{\mathbf{A}}$. If $\tilde{\mathbf{a}}$ and $\tilde{\mathbf{b}}$ differ at $l$ bits, we have:
\begin{eqnarray}\label{eq:TheoProof1}
\langle\mathbf{a} , \mathbf{b}\rangle=\frac{1}{\tilde{n}}\bigg(1\times(\tilde{n}-l)+ (-1)\times l\bigg)=\frac{\tilde{n}-2l}{\tilde{n}}
\end{eqnarray} 
Moreover, $\tilde{\mathbf{b}}$ and $\tilde{\mathbf{a}}\oplus \mathbf{1}_{\tilde{n}\times 1}$ (complement of $\tilde{\mathbf{a}}$) differ at $\tilde{n}-l$ bits and since all the three vectors $\{\mathbf{a},~\tilde{\mathbf{a}}\oplus \mathbf{1}_{\tilde{n}\times 1},~\mathbf{b}\}$ are different code words (from each complement couple, exactly one is chosen and thus $\mathbf{b}\neq\tilde{\mathbf{a}}\oplus \mathbf{1}_{\tilde{n}\times 1}$), both $l$ and $\tilde{n}-l$ should be greater than or equal to $\tilde{d}_{min}$, i.e.,:
\begin{eqnarray}\label{eq:TheoProof2}
\left\{\begin{array}{l}
l\geq\tilde{d}_{min}\\
\tilde{n}-l\geq\tilde{d}_{min}\\
\end{array}\right. &\Rightarrow& \tilde{d}_{min}\leq l\leq\tilde{n}-\tilde{d}_{min}\nonumber\\
&\Rightarrow&  |\tilde{n}-2l|\leq\tilde{n}-2\tilde{d}_{min}
\end{eqnarray}
Note that $\mathbf{0}_{\tilde{n}\times 1},\mathbf{1}_{\tilde{n}\times 1}\in\mathcal{C}$ and for each code vector $\mathbf{a}$, either $d(\mathbf{0}_{\tilde{n}\times 1} , \mathbf{a})$ or $d(\mathbf{1}_{\tilde{n}\times 1} , \mathbf{a})$ cannot exceed $\frac{\tilde{n}}{2}$; therefore, $\tilde{n}-2\tilde{d}_{min}\geq 0$. Combining (\ref{eq:TheoProof1}) and (\ref{eq:TheoProof2}) we have:
\begin{eqnarray}
|\langle\mathbf{a},\mathbf{b}\rangle|\leq \frac{\tilde{n}-2\tilde{d}_{min}}{\tilde{n}}
\end{eqnarray}
which proves the claim on the inner product of the columns of $\mathbf{A}$. Now let $\mathbf{B}_{\tilde{n}\times k}$ be the matrix formed by $k$ different columns of $\mathbf{A}$. According to the previous arguments, $\mathbf{B}^T\mathbf{B}$ is a $k\times k$ matrix that has $1$'s on its main diagonal while its off-diagonal elements have absolute values less than or equal to $\frac{\tilde{n}-2\tilde{d}_{min}}{\tilde{n}}$. It is now rather easy to complete the proof with use of the Gershgorin circle theorem $\square$

The above theorem is useful only when $\tilde{d}_{min}$ is close to $\frac{\tilde{n}}{2}$ (denominator for the upper bound of $k$), which is not the case for the common binary codes. In fact, in communication systems, parity bits are inserted to protect the main data payload, i.e., $\tilde{k}$ bits of data are followed by $\tilde{n}-\tilde{k}$ parity bits. In this case, we have $\tilde{d}_{min}\leq\tilde{n}-\tilde{k}+1$; thus, to have $\tilde{d}_{min}\approx\frac{\tilde{n}}{2}$, the number of parity bits should have the same order as the data payload which is impractical. In the next section we show how these types of codes can be designed using the well-known BCH codes.

\section{BCH codes with large $\tilde{d}_{min}$}\label{sec:dminCodes}
Since the focus in this section is on the design of BCH codes with large minimum distances, we first briefly review the BCH structure.

BCH codes are a class of cyclic binary codes with $\tilde{n}=2^{\tilde{m}}-1$ which are produced by a generating polynomial $g(x)\in GF(2)[x]$ such that $g(x)|x^{2^{\tilde{m}}-1}+1$ \cite{LinCostelloBook}. According to a result in Galois theory, we know:
\begin{eqnarray}
x^{2^{\tilde{m}}-1}+1=\prod_{\begin{array}{c}
r\in GF(2^{\tilde{m}})\\
r\neq 0
\end{array}}(x-r)
\end{eqnarray}
Hence, the BCH generating polynomial can be decomposed into the product of linear factors in $GF(2^{\tilde{m}})[x]$. Let $\alpha\in GF(2^{\tilde{m}})$ be a primitive root of the field and let $\alpha^i$ be one of the roots of $g(x)$. Since $g(x)\in GF(2)[x]$, all conjugate elements of $\alpha^i$ (with respect to $GF(2)$) are also roots of $g(x)$. Again using the results in Galois theory, we know that these conjugates are different elements of the set $\{\alpha^{i2^{j}}\}_{j=0}^{m-1}$. Moreover, since $\alpha^{2^{\tilde{m}}-1}=1$, $i_1\equiv i_2 (\textrm{mod}~2^{\tilde{m}}-1)$ implies $\alpha^{i_1}=\alpha^{i_2}$ which reveals the circular behavior of the exponents.

The main advantage of the BCH codes compared to other cyclic codes is their guaranteed lower bound on the minimum distance \cite{LinCostelloBook}: if $\alpha^{i_1},\dots,\alpha^{i_d}$ are different roots of $g(x)$ (not necessarily all the roots) such that $i_1,\dots,i_d$ form an arithmetic progression, then $\tilde{d}_{min}\geq d+1$.

Now we get back to our code design approach. We construct the desired code generating polynomials by investigating their parity check polynomial which is defined as:
\begin{eqnarray}
h(x)\triangleq \frac{x^{2^{\tilde{m}}-1}+1}{g(x)}
\end{eqnarray}

In other words, each field element is the root of exactly one of the $g(x)$ and $h(x)$. We construct $h(x)$ by introducing its roots. Let $l<\tilde{m}$ be an integer and define 
\begin{eqnarray}
\mathcal{G}_{\tilde{m}}^{(l)}=\{\alpha^{0},\alpha^{1},\dots,\alpha^{2^{\tilde{m}-1}+2^{l}-1}\}
\end{eqnarray}
Note that the definition of $\mathcal{G}_{\tilde{m}}^{(l)}$ depends on the choice of the primitive element ($\alpha$). We further define $\mathcal{H}_{\tilde{m}}^{(l)}$ as the subset of $\mathcal{G}_{\tilde{m}}^{(l)}$ which is closed with respect to the conjugate operation:
\begin{eqnarray}
\mathcal{H}_{\tilde{m}}^{(l)}\triangleq \{r\in\mathcal{G}_{\tilde{m}}^{(l)}~\big|~\forall~j\in\mathbb{N}:~r^{2^j}\in\mathcal{G}_{\tilde{m}}^{(l)}\}
\end{eqnarray}
The above definition shows that if $r\in\mathcal{H}_{\tilde{m}}^{(l)}$ then its conjugate $r^{2^j}\in\mathcal{H}_{\tilde{m}}^{(l)}$. Now let us define $h(x)$:
\begin{eqnarray}
h(x)= \prod_{r\in\mathcal{H}_{\tilde{m}}^{(l)}}(x-r)
\end{eqnarray}

As discussed before, if $r$ is a root of $h(x)$, all its conjugates are also roots of $h(x)$; therefore, $h(x)\in GF(2)[x]$, which is a required condition. Moreover,
\begin{eqnarray}
1=\alpha^{0}\in\mathcal{G}_{\tilde{m}}^{(l)} &\Rightarrow& 1\in\mathcal{H}_{\tilde{m}}^{(l)}\nonumber\\
&\Rightarrow& (1+x)\big|h(x)
\end{eqnarray}
which means that the all one vector is a valid code word:
\begin{eqnarray}
&& c=[\underbrace{1,\dots,1}_{2^{\tilde{m}}-1}]^T\nonumber\\
&\Rightarrow& c(x)=1+x+\dots+x^{2^{\tilde{m}}-2}=\frac{x^{2^{\tilde{m}}-1}+1}{x+1}\nonumber\\
&\Rightarrow& x^{2^{\tilde{m}}-1}+1\big|(x^{2^{\tilde{m}}-1}+1)\frac{h(x)}{1+x}=c(x)h(x)
\end{eqnarray}

Hence, the code generated by $g(x)=\frac{x^{\tilde{n}}+1}{h(x)}$ is a symmetric code and fulfills the requirement of Theorem \ref{theo:CodingCS}. For the minimum distance of the code, note that the roots of $h(x)$ form a subset of $\mathcal{G}_{\tilde{m}}^{(l)}$; thus, all the elements in $GF(2^{\tilde{m}})\backslash\mathcal{G}_{\tilde{m}}^{(l)}$ are roots of $g(x)$:
\begin{eqnarray}
\forall~2^{\tilde{m}-1}+2^l\leq j\leq 2^{\tilde{m}}-2:~~g(\alpha^{j})=0
\end{eqnarray}
Consequently, there exists an arithmetic progression of length $2^{\tilde{m}-1}-2^l-1$ among the powers of $\alpha$ in roots of $g(x)$. As a result:
\begin{eqnarray}
\tilde{d}_{min}\geq (2^{\tilde{m}-1}-2^l-1)+1=2^{\tilde{m}-1}-2^l
\end{eqnarray}

In coding, it is usual to look for a code with maximum $\tilde{d}_{min}$ given $\tilde{n},\tilde{k}$. Here, we have designed a code with good $\tilde{d}_{min}$ for a given $\tilde{n}$ but with unknown $\tilde{k}$:
\begin{eqnarray}
\tilde{n}&=&\tilde{k}+deg\big(g(x)\big)\nonumber\\
\Rightarrow \tilde{k}&=&\tilde{n}-deg\big(g(x)\big)\nonumber\\
&=&\big(deg\big(g(x)\big)+deg\big(h(x)\big)\big)-deg\big(g(x)\big)\nonumber\\
&=&deg\big(h(x)\big)=|\mathcal{H}_{\tilde{m}}^{(l)}|
\end{eqnarray}

The following theorem reveals how $|\mathcal{H}_{\tilde{m}}^{(l)}|$ should be calculated.

\begin{theo}\label{theo:DetermineK}
With the previous terminology, $|\mathcal{H}_{\tilde{m}}^{(l)}|$ is equal to the number of binary sequences of length $\tilde{m}$ such that if the sequence is written around a circle, between each two $1$'s, there exists at least $\tilde{m}-l-1$ zeros. 
\end{theo}

\textbf{Proof}. We show that there exists a 1-1 mapping between the elements of $\mathcal{H}_{\tilde{m}}^{(l)}$ and the binary sequences. Let $(b_{\tilde{m}-1},\dots,b_0)\in\{0,1\}^{\tilde{m}}$ be one of the binary sequences and let $\beta$ be the decimal number that its binary representation coincides with the sequence:
\begin{eqnarray}
\beta=(\overline{b_{\tilde{m}-1} \dots b_0})_2=\sum_{i=0}^{\tilde{m}-1}b_i2^i
\end{eqnarray}
We will show that $\alpha^{\beta}\in\mathcal{H}_{\tilde{m}}^{(l)}$. For the sake of simplicity, let us define $\beta_j$ as the decimal number that its binary representation is the same as the sequence subjected to $j$ units of left circular shift ($\beta_0=\beta$):
\begin{eqnarray}
\beta_0&=&(\overline{b_{\tilde{m}-1} \dots b_0})_2\nonumber\\
\beta_1&=&(\overline{b_{\tilde{m}-2} \dots b_0 b_{\tilde{m}-1}})_2\nonumber\\
\beta_2&=&(\overline{b_{\tilde{m}-3} \dots b_0 b_{\tilde{m}-1} b_{\tilde{m}-2}})_2\nonumber\\
&\vdots&\nonumber\\
\beta_{\tilde{m}-1}&=&(\overline{b_0 b_{\tilde{m}-1} \dots b_{1}})_2
\end{eqnarray}

Now we have:
\begin{eqnarray}
2\beta_{j}&=&2\times (\overline{b_{\tilde{m}-1-j} \dots b_0 b_{\tilde{m}-1} b_{\tilde{m}-j}})_2\nonumber\\
&=&2^{\tilde{m}}b_{\tilde{m}-1-j}+ (\overline{b_{\tilde{m}-2-j} \dots b_0 b_{\tilde{m}-1} b_{\tilde{m}-j}0})_2\nonumber\\
&\equiv& \beta_{j+1}~\big(\textrm{mod}~2^{\tilde{m}}-1\big)\nonumber\\
\Rightarrow && \beta_{j}\equiv 2^j\beta~\big(\textrm{mod}~2^{\tilde{m}}-1\big)\nonumber\\
\Rightarrow && \alpha^{\beta_j}=\alpha^{2^j\beta}
\end{eqnarray}
which shows that $\{\alpha^{\beta_j}\}_j$ are conjugates of $\alpha^\beta$. To show $\alpha^\beta\in\mathcal{H}_{\tilde{m}}^{(l)}$, we should prove that all the conjugates belong to $\mathcal{G}_{\tilde{m}}^{(l)}$, or equivalently, we should show $0\leq\beta_j\leq 2^{\tilde{m}-1}+2^l-1$. It is clear that $0<\beta_j$; to prove the right inequality we consider two cases:
\begin{enumerate}
\item MSB of $\beta_j$ is zero:
\begin{eqnarray}
b_{\tilde{m}-1-j}=0\Rightarrow \beta_j<2^{\tilde{m}-1}<2^{\tilde{m}-1}+2^l-1
\end{eqnarray}

\item MSB of $\beta_j$ is one; therefore, according to the property of the binary sequences, the following $\tilde{m}-l-1$ bits are zero:
\begin{eqnarray}
b_{\tilde{m}-1-j}=1&\Rightarrow& b_{\tilde{m}-2-j}=\dots=b_{l-j}=0\nonumber\\
&\Rightarrow& \beta_j\leq 2^{\tilde{m}-1}+\sum_{j=0}^{l-1}2^j\nonumber\\
&\Rightarrow& \beta_j\leq 2^{\tilde{m}-1}+2^l-1
\end{eqnarray}
\end{enumerate}

Up to now, we have proved that each binary sequence with the above zero-spacing property can be assigned to a separate root of $h(x)$. To complete the proof, we show that if the binary representation of $\beta$ does not satisfy the property, then we have $\alpha^\beta\notin\mathcal{H}_{\tilde{m}}^{(l)}$. In fact, by circular shifts introduced in $\beta_j$, all the bits can be placed in the MSB position; thus, if the binary representation of $\beta$ does not obey the property, at least one of the $\beta_j$'s should be greater than $2^{\tilde{m}-1}+2^l-1$. This means that at least one of the conjugates of $\alpha^\beta$ does not belong to $\mathcal{G}_{\tilde{m}}^{(l)}$ $\square$

Theorem \ref{theo:DetermineK} relates the code parameter $\tilde{k}$ to a combinatorics problem. Using this relation, it is shown in Appendix \ref{app:codeK} that $|\mathcal{H}_{\tilde{m}}^{(l)}|=\mathcal{O}\bigg(\big(\frac{\tilde{m}-l}{2}+1\big)^{\frac{l}{\tilde{m}-l}}\bigg)$.

\section{Sampling and Reconstruction}\label{sec:SampRec}

In previous sections, we presented the principles of matrix construction. In this section, in addition to a stepwise instruction set, we focus on the column selection procedure from complement pairs. In the second part of this section, we show that the original sparse vector can be reconstructed from the samples by simple methods such as Matching Pursuit.

\subsection{Matrix Construction}
Recalling the arguments in the previous section, the choice of the polynomial $g(x)$ depends on the choice of the primitive root. In addition to this degree of freedom, from Theorem \ref{theo:CodingCS}, no matter which code vectors from complement sets are selected, the generated matrix satisfies RIP. Hence, for a given primitive element, there are $2^{2^{\tilde{k}-1}}$ (there are $2^{\tilde{k}-1}$ complement pairs) possible matrix constructions. Among these huge number of possibilities, some of them have better characteristics for signal recovery from the samples. More specifically, we look for the matrices such that columns are closed with respect to the circular shift operation: if $\mathbf{a}=[a_1,\dots,a_{\tilde{n}}]^T$ is a column of $\mathbf{A}$, for all $1<j\leq\tilde{n}$, $\mathbf{a}_j=[a_j,a_{j+1},\dots,a_{\tilde{n}},a_1,\dots,a_{j-1}]^T$ is also a column of $\mathbf{A}$.

The key point is that the BCH codes are a subset of cyclic codes, i.e., if $\mathbf{c}_{\tilde{n}\times 1}$ is a code vector, all its circular shifts are also valid code vectors. Thus, if we are careful in selecting from the complement sets, the generated sampling matrix will also have the cyclic property. For this selection, it should be noted that if $\mathbf{a}_{\tilde{n}\times 1},\mathbf{b}_{\tilde{n}\times 1}$ is a complement pair and $\mathbf{c}_{\tilde{n}\times 1}$ is a circular shifted version of $\mathbf{a}_{\tilde{n}\times 1}$, the overal parity (sum of the elements in mod $2$) of $\mathbf{a}_{\tilde{n}\times 1}$ and $\mathbf{b}_{\tilde{n}\times 1}$ are different (each code vector has $2^{\tilde{m}}-1$ elements which is an odd number) while $\mathbf{a}_{\tilde{n}\times 1}$ and $\mathbf{c}_{\tilde{n}\times 1}$ have the same parity. Therefore, if we discard the code vectors with even (odd) parity (from the set of all code vectors), we are left with a set half the size of the main set such that from each complement set exactly one is selected while the set is still closed with respect to the circular shift operation. The selection algorithm is as follows:

\begin{enumerate}
\item For a given $k$ (compressed sensing parameter), let $i=\lceil\log_2(k)\rceil$ and choose $\tilde{m}\geq i$ (the number of compressed samples will be $m=2^{\tilde{m}}-1$).

\item \label{step:SetDef} Let $\mathcal{H}_{seq}$ be the set of all binary sequences of length $\tilde{m}$ such that $1$'s are circularly spaced with at least $i$ zeros. In addition, let $\mathcal{H}_{dec}$ be the set of decimal numbers such that their binary representation is a sequence in $\mathcal{H}_{seq}$.

\item Choose $\alpha$ as one of the primitive roots of $GF(2^{\tilde{m}})$ and define:
\begin{eqnarray}
\mathcal{H}=\{\alpha^r~\big|~r\in\mathcal{H}_{dec}\}
\end{eqnarray}

\item Define the parity check and code generating polynomials as:
\begin{eqnarray}
h(x)=\prod_{r\in\mathcal{H}}(x-r)
\end{eqnarray}
and
\begin{eqnarray}
g(x)=\frac{x^{2^{\tilde{m}}}-1}{h(x)}
\end{eqnarray}

\item Let $\tilde{\mathbf{A}}_{(2^{\tilde{m}}-1)\times(2^{deg(h)-1})}$ be the binary matrix composed of even parity code vectors as its columns, i.e., if columns are considered as polynomial coefficients (in $GF(2)[x]$), each polynomial should be divisible by $(x+1)g(x)$ (the additional factor of $x+1$ implies the even parity).

\item Replace all the zeros in $\tilde{\mathbf{A}}$ by $-1$ and normalize each column to obtain the final compressed sensing matrix ($\mathbf{A}_{(2^{\tilde{m}}-1)\times(2^{deg(h)-1})}$).
\end{enumerate}

For a simple example, we consider the case $\tilde{m}=i$. It is easy to check that the number of $1$'s in each of the binary sequences in step \ref{step:SetDef} cannot exceed one. Therefore, we have $\mathcal{H}_{dec}=\{0,2^0,2^1,2^2,\dots,2^{2^{i-1}}\}$. This means that $h(x)$, except for the factor $(x+1)$ is the same as the minimal polynomial of $\alpha$ (the primitive root). Since for code generation, we use $(x+1)g(x)$ instead of $g(x)$, the effective $h(x)$ will be the minimal polynomial of $\alpha$ which is a primitive polynomial. In this case, the matrix $\tilde{\mathbf{A}}$ is the $(2^i-1)\times(2^i-1)$ square matrix whose columns are circularly shifted versions of the Pseudo Noise Sequence (PNS) output generated by the primitive polynomial (the absolute value of the inner product of each two columns of $\mathbf{A}$ is exactly $\frac{1}{2^i-1}$).

\begin{figure}[tb]
\centering
\includegraphics[width=8cm]{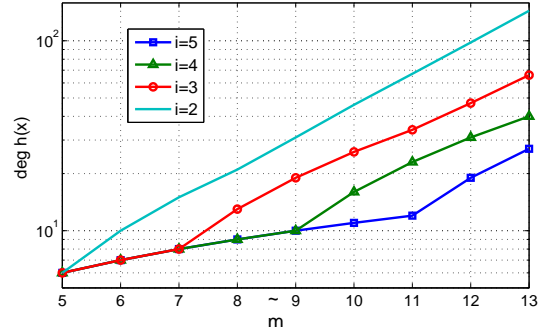}
\caption{Degree of $h(x)$ for different values of $\tilde{m}$ and $i$.}
\label{fig:degH}
\end{figure}

Table \ref{tab:pm1Matrixi3} summarizes some of the parity check polynomials for $i=3$ (useful for $k<8$). Also, Fig. \ref{fig:degH} shows the degree of $h(x)$ for some of the choices of $\tilde{m}$ and $i$.

\begin{table}[t]
\centering
\begin{tabular}{|c|c|}
\hline
$\tilde{m}$ & $h(x)$\\
\hline\hline
$4$ & $x^5+x^4+x^2+1$\\
\hline
$6$ & $x^7+x^6+x^2+1$\\
\hline
$8$ & $x^{13}+x^{12}+x^{10}+x^9+x^8+x^4+x^3+1$\\
\hline
$10$ & $x^{26}+x^{25}+x^{24}+x^{20}+x^{16}+x^{14}+x^{13}+x^{12}$\\
 & $+x^{10}+x^9+x^7+x^5+x^4+x^3+x+1$\\
\hline
\end{tabular}
\caption{Parity check polynomials for different values of $\tilde{m}$ when $i=3$.}
\label{tab:pm1Matrixi3}
\end{table}

\subsection{Reconstruction from the samples}
Matching Pursuit is one of the simplest methods for the recovery of sparse signals from sampling matrices (linear projections). Here we show that this method can exactly recover the sparse signal from noiseless samples.

Let $\mathbf{A}_{m\times n}$ and $\mathbf{s}_{n\times 1}$ be the sampling matrix and the $k$-sparse signal vector, respectively. The sampling process is defined by:
\begin{eqnarray}
\mathbf{y}_{m\times 1}=\mathbf{A}_{m\times n}\cdot\mathbf{s}_{n\times 1}
\end{eqnarray} 

For unique reconstruction of $\mathbf{s}_{n\times 1}$ from the samples $\mathbf{y}_{m\times 1}$, it is sufficient that the sampling matrix $\mathbf{A}_{m\times n}$ satisfies RIP of order $2k$ \cite{Candes2008}. In this section, we show that if $\mathbf{A}_{m\times n}$ is constructed as described in previous section and satisfies RIP of order $2k$, the matching pursuit method can be used for perfect reconstruction. In addition, due to the circular structure of the columns in $\mathbf{A}_{m\times n}$, the computational complexity can be decreased (less than the ordinary matching pursuit).

Let $\{i_1,\dots,i_k\}\subset\{1,\dots,n\}$ be the nonzero locations in $\mathbf{s}_{n\times 1}$; thus, we have:
\begin{eqnarray}
\mathbf{y}_{m\times 1}=\mathbf{A}\cdot\mathbf{s}=\sum_{j=1}^k s_{i_j}\mathbf{a}_{i_j}
\end{eqnarray}
where $\mathbf{a}_{i}$ denotes the $i^{th}$ column in $\mathbf{A}$. In the matching pursuit method, in order to find the nonzero locations in $\mathbf{s}$, the inner products of the sample-vector ($\mathbf{y}$) with all the columns of $\mathbf{A}$ are evaluated and then, the index of the maximum value (in absolute) is chosen as the most probable nonzero location. Here, we show that the index associated with the maximum value is always a nonzero location. Without loss of generality, assume $|s_{i_1}|\geq|s_{i_2}|\geq\dots\geq|s_{i_k}|$. We then have:
\begin{eqnarray}\label{eq:Biggerineq}
\big|\langle\mathbf{y},\mathbf{a}_{i_1}\rangle\big|&=&\big|\sum_{j=1}^{k}s_{i_j} \langle\mathbf{a}_{i_j},\mathbf{a}_{i_1}\rangle\big|\nonumber\\
&\geq& |s_{i_1}|\langle\mathbf{a}_{i_1},\mathbf{a}_{i_1}\rangle- \sum_{j=2}^{k}|s_{i_j}||\langle\mathbf{a}_{i_j},\mathbf{a}_{i_1}\rangle|\nonumber\\
&>& |s_{i_1}| - \frac{1}{2k-1}\sum_{j=2}^{k}|s_{i_j}|\nonumber\\
&\geq& |s_{i_1}| - \frac{k-1}{2k-1}|s_{i_1}|=\frac{k}{2k-1}|s_{i_1}|
\end{eqnarray}

Now assume $l\in\{1,\dots,n\}\backslash \{i_1,\dots,i_k\}$:
\begin{eqnarray} \label{eq:Smallerineq}
\big|\langle\mathbf{y},\mathbf{a}_{l}\rangle\big|&=&\big|\sum_{j=1}^{k}s_{i_j} \langle\mathbf{a}_{i_j},\mathbf{a}_{l}\rangle\big|\nonumber\\
&\leq&  \sum_{j=1}^{k}|s_{i_j}||\langle\mathbf{a}_{i_j},\mathbf{a}_{l}\rangle|\nonumber\\
&<& \frac{1}{2k-1}\sum_{j=1}^{k}|s_{i_j}|\leq\frac{k}{2k-1}|s_{i_1}|
\end{eqnarray}

Combining (\ref{eq:Biggerineq}) and (\ref{eq:Smallerineq}), we get:
\begin{eqnarray}
\big|\langle\mathbf{y},\mathbf{a}_{l}\rangle\big|~<~\frac{k}{2k-1}|s_{i_1}|~<~ \big|\langle\mathbf{y},\mathbf{a}_{i_1}\rangle\big|
\end{eqnarray}
Hence, the largest inner product is obtained either with $\mathbf{a}_{i_1}$ or one of the other $\mathbf{a}_{i_j}$'s. Therefore, in the noiseless case, we never select a nonzero location by using the matching pursuit algorithm, and finally we reconstruct the original sparse signal perfectly.

In each recursion of the matching pursuit algorithm, the inner product of $\mathbf{y}_{m\times 1}$ with all the columns in $\mathbf{A}_{m\times n}$ needs to be calculated. Each inner product requires $m$ multiplications and $m-1$ additions. Now we observe that the circular property of the columns of $\mathbf{A}$ can be useful. Let $\mathbf{a}$ be one of the columns in $\mathbf{A}$ and $\mathbf{a}^{(j)}$ be its $j^{th}$ circularly shifted version. We observe that $\{\mathbf{a}^{(j)}\}_j$ are all columns of $\mathbf{A}$; thus, $\langle\mathbf{a}^{(j)},\mathbf{y}\rangle$ has to be calculated for all $j$. Let $\{\mathbf{a}^{(1)} , \mathbf{a}^{(2)} , \dots , \mathbf{a}^{(\mu)}\}$ be different elements of $\{\mathbf{a}^{(j)}\}_j$ (obviously $\mu\leq m$ and more precisely $\mu|m$). These inner products require $\mu m$ multiplications and $\mu (m-1)$ additions if directly calculated.

An alternative approach for evaluation of these values is to employ Discrete Fourier Transform (DFT) or its fast implementation-FFT. The key point in this approach is that the inner products can be found through circular convolution of $\mathbf{y}$ and $\mathbf{a}$, i.e.,
\begin{eqnarray}
\langle\mathbf{y},\mathbf{a}_{(j)}\rangle=\mathbf{y}\varoast_{m} \mathbf{a}\big|_{j}
\end{eqnarray}
where $\varoast_m$ represents the circular convolution with period $m$. It is well-known that the circular convolution can be easily calculated using DFT: if $\mathbf{y}_f$ and $\mathbf{a}_f$ denote the DFT of $\mathbf{y}$ and $\mathbf{a}$, respectively, we have:
\begin{eqnarray}
IDFT\{\mathbf{y}_f\odot\mathbf{a}_f\}=\big[\mathbf{y}\varoast_m \mathbf{a}\big|_{0},\dots,\mathbf{y}\varoast_m \mathbf{a}\big|_{m-1}\big]
\end{eqnarray}
where $\mathbf{v}_{m\times 1}\odot\mathbf{u}_{m\times 1}\triangleq[v_1u_1,\dots,v_mu_m]^T$. For evaluation of the inner products in this way, $\mathbf{y}_f$has to be calculated only once using DFT. Thus, excluding the calculation of $\mathbf{y}_f$ (which is done only once), the inner products of $\mathbf{y}$ with $\{\mathbf{a}^{(j)}\}_j$ require one $DFT$, one $IDFT$ and $m$ multiplications. Since $\mu$ different circular shifts of $\mathbf{a}$ are possible, at most $\mu$ coefficients of $\mathbf{a}_f$ at equi-distance positions are nonzero; hence, $\mu$-point DFT (and consequently IDFT) of $\mathbf{a}_{m\times 1}$ rather than the general $m$-point DFT is adequate. 
For $\mu$-point DFT of $\mathbf{y}$, we can simply down-sample the evaluated $m\times 1$ vector of $\mathbf{y}_f$ (note that $\mu|m$) and there is no need for an extra $\mu$-point DFT. Employing the FFT version, we require $2\mu\lceil\log_2\mu\rceil$ multiplications and $m-\mu+2\mu\lceil\log_2\mu\rceil$ additions per $\mu$-point DFT or IDFT. Comparing the number of required multiplications in calculation of the above $\mu$ inner products reveal the efficiency of the DFT approach; i.e., the required computational complexity for reconstruction of the signal from the samples obtained from the sampling matrix is less than the common amount for general matrices. It should be emphasized that this reduction in computational complexity is the result of the circular format of the columns.

\section{Matrices with $\{0,1,-1\}$ Elements}\label{sec:ThreeElement}

We have presented a method to generate RIP-fulfilling matrices with $\pm 1$ elements. In this section, we show that the matrices introduced in \cite{DeVore2007} can be improved using our technique in this paper.

In \cite{DeVore2007}, in contrast to this paper, binary compressed sensing matrices are considered. The main difficulty in designing such matrices is that the columns should (almost) be normal which means that prior to normalization, the number of $1$'s in each column is fixed (matrix elements are all scaled with the same coefficient for normalization). In \cite{DeVore2007}, $p^2\times p^{r+1}$ binary matrices are introduced such that in each column, exactly $p$ elements are equal to $1$ (equal to $\frac{1}{\sqrt{p}}$ after normalization) and the inner product of each two columns is less than or equal to $r$ ($\frac{r}{p}$ after normalization). Here $p$ is a power of a prime integer; the matrix construction is based on polynomials in $GF(p)$.

It is evident that by changing some of the $1$'s in the aforementioned matrix into $-1$, the norm of the columns does not change; however, the inner products change. To show how we can benefit from this feature, let us assume that $p=2^i$; thus, there are $2^i$ nonzero elements in each column. We construct a new matrix from the original binary matrix as follows: we repeat each column $2^i$ times and then change the sign of the nonzero elements in the replicas in such a way that these nonzero elements form a Walch-Hadamard matrix. In other words, for each column, there are $2^i$ columns (including itself) that have the same pattern of nonzero elements. Moreover, the nonzero elements of these semi-replica vectors are different columns of the Walch-Hadamard matrix. Thus, the semi-replica vectors are orthogonal and the absolute value of the inner product of two vectors with different nonzero patterns is upper-bounded by $r$ (maximum possible value in the original matrix). Hence, the new matrix still satisfies the RIP condition with the same $k$ and $\delta_k$. 

Although we have expanded the matrix with this trick, the change is negligible when the order of matrix sizes are considered ($p^2\times p^{r+1}$ is expanded to $p^2\times p^{r+2}$). In fact, the orthogonality of the semi-replicas is not a necessary condition; we only need that their inner products do not exceed $r$ in absolute value. It shows that instead of the Walch-Hadamard matrix, we can use other $\pm 1$ matrices with more number of columns (with the same number of rows) such that their columns are almost orthogonal (inner product less than $r$). This is the case for the matrices introduced in the previous sections.

In order to mathematically describe the procedure, we need to define an operation. Let $\mathbf{s}$ be a $\beta\times 1$ binary vector with exactly $\alpha$ elements of $1$ in locations $r_1,\dots,r_{\alpha}\in\{1,2,\dots,\beta\}$. Also, let $\mathbf{x}_{\alpha\times 1}=[x_1,\dots,x_{\alpha}]^T$ be an arbitrary vector. We define $\mathbf{y}_{\beta\times 1}=\mu(\mathbf{s},\mathbf{x})$ as:
\begin{eqnarray}
\left\{\begin{array}{llll}
\forall~1\leq j\leq \alpha: & y_{r_j} & =x_{j}\\
\forall~j\notin\{r_1,\dots,r_{\alpha}\}:& y_j & =0
\end{array}\right.
\end{eqnarray}

From the above definition, we can see:
\begin{eqnarray}
\langle\mu(\mathbf{s},\mathbf{x}_1)\;,\;\mu(\mathbf{s},\mathbf{x}_2)\rangle=\langle\mathbf{x}_1,\mathbf{x}_2\rangle
\end{eqnarray}
Furthermore, if the elements of both $\mathbf{x}_1,\mathbf{x}_2$ lie in the closed interval $[-1,1]$, we have:
\begin{eqnarray}
\big|\langle\mu(\mathbf{s}_1,\mathbf{x}_1)\;,\;\mu(\mathbf{s}_2,\mathbf{x}_2)\rangle\big|\leq\langle\mathbf{s}_1,\mathbf{s}_2\rangle
\end{eqnarray}

For the matrix construction, let $\tilde{m}$ be an integer such that $p=2^{\tilde{m}}-1$ is a prime (the primes of this form are called Mersenne primes). Let $k<p$ be the required order of the RIP condition and let:
\begin{eqnarray}
r=\big\lfloor\frac{p}{k}\big\rfloor~~,~~i=\lceil\log_2 k\rceil
\end{eqnarray}

Also let $\mathbf{S}_{p^2\times p^{r+1}}=[\mathbf{s}_1~\dots~\mathbf{s}_{p^{r+1}}]$ be the binary RIP-fulfilling matrix constructed as in \cite{DeVore2007} and $\mathbf{X}_{p\times 2^{\tilde{k}}}=[\mathbf{x}_1~\dots~\mathbf{x}_{2^{\tilde{k}}}]$ ($\tilde{k}=|\mathcal{H}_{\tilde{m}}^{(\tilde{m}-i)}|$ with the previous terminology) be the $\pm 1$ matrix introduced in the previous sections (we further normalize the columns of these matrices). We construct a new $p^2\times (p^{r+1}.2^{\tilde{k}})$ matrix with elements in $\{0,1,-1\}$ by combining these two matrices:
\begin{eqnarray}
\mathbf{A}=[\mu(\mathbf{s}_i,\mathbf{x}_j)]_{i,j}
\end{eqnarray}
Employing the same approach as used before, we show that $\mathbf{A}$ satisfies the RIP condition of order $k$, i.e., we show that the inner product of two different columns of $\mathbf{A}$ cannot exceed $\frac{1}{k-1}$ in absolute value while each column is normal:
\begin{eqnarray}
\langle~\mu(\mathbf{s}_i,\mathbf{x}_j)~,~\mu(\mathbf{s}_i,\mathbf{x}_j)~ \rangle=\langle \mathbf{x}_j , \mathbf{x}_j\rangle = 1
\end{eqnarray}

To study the inner product of $\mu(\mathbf{s}_{i_1},\mathbf{x}_{j_1}$ and $\mu(\mathbf{s}_{i_2},\mathbf{x}_{j_2}$, we consider two cases:
\begin{enumerate}
\item $i_1=i_2$. In this case, since $\mathbf{s}_{i_1}=\mathbf{s}_{i_2}$, we have:
\begin{eqnarray}\label{eq:IneqX}
\big|\langle~\mu(\mathbf{s}_{i_1},\mathbf{x}_{j_1})~,~\mu(\mathbf{s}_{i_2},\mathbf{x}_{j_2})~ \rangle\big|&=&\big|\langle~\mathbf{x}_{j_1}~,~\mathbf{x}_{j_2}~ \rangle\big|\nonumber\\
&<&\frac{1}{k-1}
\end{eqnarray}

\item $i_1\neq i_2$ and therefore, $\mathbf{s}_{i_1}\neq\mathbf{s}_{i_2}$; since the elements of both $\mathbf{x}_{j_1}$ and $\mathbf{x}_{j_1}$ lie in $[-1,1]$, we have:
\begin{eqnarray}\label{eq:IneqS}
\big|\langle~\mu(\mathbf{s}_{i_1},\mathbf{x}_{j_1})~,~\mu(\mathbf{s}_{i_2},\mathbf{x}_{j_2})~ \rangle\big|&\leq&\big|\langle~\mathbf{s}_{i_1}~,~\mathbf{s}_{i_2}~ \rangle\big|\nonumber\\
&<&\frac{1}{k-1}
\end{eqnarray}

\end{enumerate}

Inequalities (\ref{eq:IneqX}) and (\ref{eq:IneqS}) hold due to the RIP-fulfilling structure of the matrices $\mathbf{X}$ and $\mathbf{S}$.
Hence, the claimed property of the inner products of the columns in $\mathbf{A}$ is proved. Consequently, $\mathbf{A}$ obeys the RIP condition of order $k$.

\section{Conclusion}\label{sec:Conclusion}

Despite the enormous amount of literature in random sampling matrices for compressed sensing, deterministic designs are not well researched. In this paper, we introduce a new connection between the coding theory and RIP fulfilling matrices. In the new design, we replace the zeros in the binary linear code vectors by $-1$ and use them as the columns of the sampling matrix in compressed sensing. The advantage of these matrices, in addition to their deterministic and known structure, is the simplicity in the sampling process; real/complex entries in the sampling matrix increases the computational complexity of the sampler as well as the required bit-precision for storing the samples. The linear codes for this purpose should have some desired characteristics; existence of such linear codes is proved by explicitly introducing binary BCH codes. One of the features of these matrices is that their produced samples can be easily (using matching pursuit method) decoded as the original sparse vector and due to the circular structure of the columns, the computational complexity in recovery can be reduced. These $\pm 1$ matrices are further expanded by considering $\{0,1,-1\}$ elements; this expansion is achieved by combining the $\pm 1$ matrices introduced in this paper with the Devore's binary matrices. Although the generated matrices show an improvement in the realizable size of the RIP-constrained matrices, the bound predicted by random matrices is not achieved yet.


%

\appendices
\section{Evaluation of $\tilde{k}$}\label{app:codeK}
In Theorem \ref{theo:DetermineK}, we showed that $\tilde{k}$ is equal to the number of binary sequences of length $\tilde{m}$ such that no two $1$s are spaced by less than $\tilde{m}-l-1$ zeros (circular definition). To evaluate this number, let us define $\tau^{(a)}_{b}$ as the number of binary sequences of length $b$ such that if the sequence is put around a circle, between each two $1$'s, there is at least $a$ zeros. In addition, let $\kappa_{b}^{(a)}$ be the number of binary sequences such $1$'s are spaced by at least $a$ zeros apart (circular property is no longer valid for $\kappa_{b}^{(a)}$). We first calculate $\kappa_{b}^{(a)}$ and then we show the connection between $\kappa_{b}^{(a)}$ and $\tau_{b}^{(a)}$.

There are two kinds of binary sequences counted in $\kappa_{b}^{(a)}$:
\begin{enumerate}
\item The last bit in the sequence is $0$; by omitting this bit, we obtain a sequence of length $b-1$ with the same property. Also, each binary sequence of length $b-1$ with the above property can be padded by $0$ while still satisfying the required property to be included in $\kappa_{b}^{(a)}$. Therefore, there are $\kappa_{b-1}^{(a)}$ binary sequence of this type.

\item The last bit in the sequence is $1$; this means that the last $a+1$ bits of the sequence are $\underbrace{0,\dots,0}_{a},1$. Similar to the above case, each binary sequence of length $b-a-1$ counted in $\kappa_{b-a-1}^{(a)}$ can be padded by the block $\underbrace{0,\dots,0}_{a},1$ to produce a sequence included in $\kappa_{b}^{(a)}$. Thus, there are $\kappa_{b-a-1}^{(a)}$ binary sequences of this type.
\end{enumerate}

In summary, we have the following recursive equation:
\begin{eqnarray}\label{eq:DifferenceEq}
\kappa_{b}^{(a)}=\kappa_{b-1}^{(a)}+\kappa_{b-a-1}^{(a)}
\end{eqnarray}
Since for $b\leq a+1$, there can be at most one $1$ in the binary sequence, we thus have:
\begin{eqnarray}\label{eq:InitialCond}
1\leq b\leq a+1:~\kappa_{b}^{(a)}=b+1
\end{eqnarray}
From (\ref{eq:DifferenceEq}), the last initial condition ($\kappa_{a+1}^{(a)}=a+2$) is equivalent to $\kappa_{0}^{(a)}=1$. If we define the onesided $\mathcal{Z}$-transform of $\kappa_{b}^{(a)}$ as follows
\begin{eqnarray}
\kappa^{(a)}(z)=\sum_{b=0}^{\infty}\kappa_{b}^{(a)}z^{-b},
\end{eqnarray}
it is not hard to check that:
\begin{eqnarray}
\kappa^{(a)}(z)=\frac{z}{z-1}\cdot\frac{z^{a+1}-1}{z^{a+1}-z^a-1}
\end{eqnarray}
Therefore, the increasing rate $\kappa_{b}^{(a)}$ with respect to $b$ ($b\gg 1$) has the same order as $\gamma^b$ where $\gamma$ is the largest (in absolute value) root of $f(z)=z^{a+1}-z^a-1$. Since $f(1)\cdot f(2)<0$, there is a real root in $(1~,~2)$; let us denote this root by $\gamma$. In fact, $\gamma$ is the largest root of $f(z)$ (we do not prove this; however, if $f(z)$ has a larger root, the increasing rate of $\kappa_{b}^{(a)}$ would be greater than $\gamma^b$):
\begin{eqnarray}
1<\gamma <2~~,~~~ f(\gamma)=\gamma^{a+1}-\gamma^a-1=0
\end{eqnarray}

Since $\gamma>1$ we have $\gamma^{a+1}>1$. For the sake of simplicity, let us define:
\begin{eqnarray}
\delta\triangleq \gamma^{a+1} - 1
\end{eqnarray}
We thus have:
\begin{eqnarray}\label{eq:rootineq}
&&(1+\delta) - (1+\delta)^{\frac{a}{a+1}}=1\nonumber\\
&\Rightarrow& (1+\delta)^{\frac{a}{a+1}}\bigg((1+\delta)^{\frac{1}{a+1}}-1\bigg)=1\nonumber\\
&\Rightarrow& (1+\delta)^{\frac{a}{a+1}}\big((1+\delta)-1\big)=\sum_{j=0}^{a}(1+\delta)^{\frac{j}{a+1}}\nonumber\\
&\Rightarrow& \delta(1+\delta)^{\frac{a}{a+1}}\geq \sum_{j=0}^{a} 1+\frac{j}{a+1}\delta\nonumber\\
&\Rightarrow& \delta(1+\delta)\geq\frac{a}{2}\delta+a+1\nonumber\\
&\Rightarrow& \big(\delta -\frac{a-2}{4}\big)^2\geq\frac{(a-2)^2}{16}+a+1>\bigg(\frac{a+4}{4}\bigg)^2\nonumber\\
&\Rightarrow& \delta>\frac{a+1}{2}~~\Rightarrow~~ 1+\delta>\frac{a+3}{2}\nonumber\\
&\Rightarrow& \gamma>\bigg(\frac{a+3}{2}\bigg)^{\frac{1}{a+1}}
\end{eqnarray}

Now we can show the connection between $\tau_b^{(a)}$ and $\kappa_b^{(a)}$. According to the definition of these parameters, we see that every binary sequence counted in $\tau_b^{(a)}$ is also counted in $\kappa_b^{(a)}$, therefore:
\begin{eqnarray}
\tau_b^{(a)}\leq\kappa_b^{(a)}
\end{eqnarray}
In addition, if a sequence counted in $\kappa_{b-a}^{(a)}$ is padded with $a$ zeros at the end, it satisfies the requirements to be counted in $\tau_b^{(a)}$, thus:
\begin{eqnarray}
\kappa_{b-a}^{(a)}\leq\tau_b^{(a)}
\end{eqnarray}

Combining the latter two inequalities, we get:
\begin{eqnarray}
\mathcal{O}(\gamma^{b-a})\leq \tau_b^{(a)} \leq\mathcal{O}(\gamma^{b})
\end{eqnarray}
The above equation in conjunction with the result in (\ref{eq:rootineq}), yields:
\begin{eqnarray}
\tau_b^{(a)}\gtrapprox \mathcal{O}\bigg(\bigg(\frac{a+3}{2}\bigg)^{\frac{b}{a+1}-1}\bigg)
\end{eqnarray}

The interpretation of the above inequality for $\tilde{k}$ is as follows:
\begin{eqnarray}
\tilde{k}=\tau_{\tilde{m}}^{(\tilde{m}-l-1)}\gtrapprox \mathcal{O}\bigg(\big(\frac{\tilde{m}-l}{2}+1\big)^{\frac{l}{\tilde{m}-l}}\bigg)
\end{eqnarray}

\begin{figure}[tb]
\centering
\includegraphics[width=8cm]{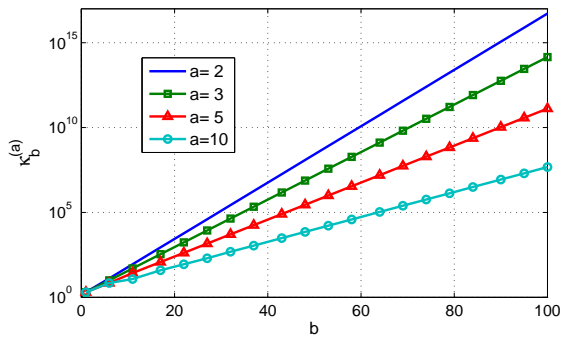}
\caption{Exact values of $\kappa_b^{(a)}$ for different values of $a$ and $b$.}
\label{fig:kappa}
\end{figure}

Figure \ref{fig:kappa} shows the asymptotic behavior of $\kappa_b^{(a)}$ at different $a$ values when $b$ increases.

\section*{Acknowledgment}
The authors sincerely thank K. Alishahi for his help in the proof given in the appendix.
%

\ifCLASSOPTIONcaptionsoff
  \newpage
\fi



%

\bibliographystyle{IEEEtran}
\bibliography{CS_bibliography}

%

%






\end{document}